\documentclass[11pt, oneside]{article}   	
\usepackage{geometry}                		
\usepackage[parfill]{parskip}    		
\usepackage{graphicx}				
\usepackage{amssymb}
\usepackage{xcolor}

\usepackage{dcolumn}
\usepackage{bm}

\def\uv{\bm{u}}
\def\rv{\bm{r}}
\def\D{{{\cal{D}}}}
\def\F{{{\cal{F}}}}

\def\Rv{{\mathbf{R}}}

\title{Correlation functions for strongly confined wormlike chains.}
\author{Joel Gard\textsuperscript{1}, Greg Morrison\textsuperscript{1,2}, \textsuperscript{1}Dept of Physics, University of Houston, Houston TX 77204.\\ \textsuperscript{2}Center for Theoretical Biophysics, Rice University, Houston TX 77005. }
\date{}

\begin{document}


\maketitle

\begin{abstract}

Polymer models describing the statistics of biomolecules under confinement have applications to a wide range of single molecule experimental techniques and give insight into biologically relevant processes {\em{in vivo}}.  In this paper, we determine the transverse position and bending correlation functions for a wormlike chain confined within slits and cylinders (with one and two confined dimensions, respectively) using a mean field approach that enforces rigid constraints on average.  We show the theoretical predictions accurately capture the statistics of a wormlike chain from Monte Carlo simulations in both confining geometries for both weak and strong confinement.  We also show that the longitudinal correlation function is accurately computed for a chain confined to a slit, and leverage the accuracy of the model to suggest an experimental technique to infer the (often unobservable) transverse statistics from the (directly observable) longitudinal end-to-end distance.  

\end{abstract}

\section{\label{introduction.sec}Introduction}

Biomolecules are found in complex and highly constrained environments {\em{in vivo}}, with examples ranging from the confinement of DNA within capsids\cite{Petrov:2008ji,Katzav:2006br,SMITH:2001cz} to crowded cellular environments\cite{Ando:2010ea,Samiotakis:2011ev}.  This restriction of the space of available conformations may have a significant effect on the shape and biological function\cite{Zhou:2008hta,Yi:2006ct,Zimmerman:1993ug} of a molecule, and alter the binding kinetics\cite{Zhou:2001hg,Samiotakis:2011ev} or the dynamical timescales\cite{Richter:2019eu,Thuroff:2011ku,Reisner:2005hr,Muthukumar:2001ka,Dai:2015cc,Brochard:1977gj,Tree:2013ev} of the system {\em{in vivo}}.  A number of studies have also modeled the prokaryotic chromosome as a confined homopolymer\cite{Ha:2015ki,Jun:2008ed,Jun:2006fd} to predict structural and dynamic features, although the applicability of homopolymers to eukaryotic chromosomes remains contested\cite{Shi:2018eo,Kang:2015hh}.  {\em{In vitro}}, a wide range of single molecule experimental setups incorporate confinement to channels or slits\cite{Hsieh:2008uo,de2015force,Yeh:2016bs} to elongate DNA or other stiff molecules for sequencing or imaging, study the translocation rates through channels\cite{Wong:2008ku,Tree:2012hf,Dekker:2021am}, and entropically pin large macromolecules\cite{Han:2002ci,Han:2000bb}.  Confinement to a slit (parallel plates) has been useful in the visualization of single molecules under applied forces or stretched via fluidic drag\cite{Bonthuis:2008fd,Huang:2014gf}.  The potential for high-throughput nanopore DNA sequencing technology\cite{vanderMaarel:2014dm}, in which DNA is translocated through a channel confinement with the intent of sequentially reading the nucleic acid sequence, is limited by noise\cite{Jo:2007gha,Min:2011hm} that is dictated by the statistics and dynamics of a confined polymer.  In each of these diverse cases, a complete understanding of the statistical properties of confined polymers is essential to fully understand biologically relevant dynamics {\em{in vivo}} or interpret experimental results {\em{in vitro}}.

Due to the ubiquity and importance of confined molecules {\em{in vivo}} and {\em{in vitro}}, a number of studies have examined the effect of spherical\cite{Morrison:2009gb}, cylindrical or square nanochannel\cite{Morrison:2005gz}, or slit\cite{Cordeiro:1997fu} (parallel plate) geometries on the statistics of a homopolymer.   Confinement effects on a homopolymer depend strongly on the flexibility of the chain (having persistence length $l_p$) in comparison to the confinement dimension $R$ (in this paper $R$ denotes the radius of a cylinder or half the separation between the plates).   The conformations adopted by flexible chains are dictated solely by the entropy and are well described by Flory theory\cite{Hsieh:2008fa,de1979scaling}, whereas the free energy for wormlike chains\cite{doi1988theory} (WLCs) has both entropic and enthalpic contributions that give rise to a new length scale:   the deflection length\cite{Odijk:1983jha}, $l_d=(l_pR^2)^{1/3}$.   Extensive theoretical, computational, and experimental work for both cylinders and slits\cite{Leith:2016il,Chen:2013dl,Cifra:2009ch,Chen:2006cq,Chen:2007eu,Dijkstra:1993,Burkhardt:1999fl,Chen:2016gy,Odijk:1983jha} have confirmed the free energy scales with $l_d$ as $\beta F\sim L/l_d$.  

The simple wormlike chain model accurately describes the statistics of real biomolecules trapped in a highly restricted confined environment, despite the idealizations built into the model.  Simulations have shown that excluded volume effects do not significantly affect the statistics of a strongly confined WLC\cite{Gupta:2015ib,Muralidhar:2014fy, Bashman;2021jcp} for the chain thickness $w$ sufficiently small.  Hydrodynamics has also been demonstrated to have weak effects on the dynamics of a strongly confined chain\cite{Dorfman:2014fd}.  Electrostatic effects on charged wormlike chains like DNA largely act to modify the persistence length through the Odijk-Skolnik-Fixman (OSF) theory\cite{Skolnick:1977jv,Odijk:1978cn}, with an electrostatic persistence length $l_{el}\propto\kappa_{debye}^{-2}$ with $\kappa_{debye}$ the inverse screening length of the solvent (although there are a number of competing theories\cite{Dobrynin:2005kq,Ha:1995iy} that predict $l_{el}\sim \kappa_{debye}^{-1}$).  Regardless of the functional form of the theory, slit confinement does not significantly alter the ionic effects on the persistence length\cite{Hsieh:2008fa}, suggesting an effective wormlike chain can still be used to describe such a system.  Theoretical modeling of real biomolecules confined to slits or cylinders has thus often focused on the strongly confined regime\cite{Jo:2007gha,Werner:2018dm}, and the interpretation of many experiments are based on predictions of the WLC model (such as the extension of the chain).  

In this paper, we use a mean field approach\cite{Hinczewski:2011el,Morrison:2009gb,Hansen:2001eka,Ha:1997bg,Ha:1995db} to analytically determine the effect of confinement on the position and bending between two points along the backbone of a wormlike chain in the transverse direction for both cylindrical and slit confinement, a quantity not readily computed solely from the linear extension. We find that strong confinement produces a damped oscillatory transverse correlation functions, similar to those found for spherical confinement\cite{Morrison:2005gz,Spakowitz:2003ca} and in qualitative agreement with a weakly bending rod approximation\cite{Wagner:2007fc}. The mean field theory predicts the emergence of two distinct length scales in the correlation functions:  the deflection length that dominates the exponential decay of correlations, and another length scale that dictates the frequency of oscillations in the correlation functions. The theoretical predictions of the mean field theory are better able to capture the behavior of Monte Carlo simulations of the correlation function over a wider range than existing weakly bending rod theories, illustrating the utility of the theory. The model requires two parameters undetermined by the theory that can be readily computed via simulations, and we discuss how these parameters can be estimated experimentally.

\section{\label{model.sec}Polymer model}

\subsection{\label{KPsub.sec}Wormlike chains and confinement}

The classic Kraty-Porod (KP) model\cite{doi1988theory} forms the discrete foundation of the inextensible Wormlike chain (WLC) model, with a chain composed of $N$ bond vectors $\uv_i$ of length $|\uv_i|=1$ with $\beta E_{bend}=-l_p/a\ \sum_{i=1}^N \uv_i\cdot\uv_{i+1}$ accounting for an intrinsic resistance to bending over the persistence length, $l_p=\beta\kappa a$, with $\kappa$ the energetic penalty to bending and $\beta=(k_BT)^{-1}$ the Boltzmann factor.  In the continuum limit (with $a\to0$, $N\to\infty$, and $L=Na$, $k/Na$ fixed), the energy associated with a particular configuration is 
$ E_{wlc}=\frac{\kappa}{2}\int_0^L ds( \partial_s{\hat{\uv}})^2+$const, with $\partial_s\hat{\uv}=\partial \hat{\uv}/\partial s$ denoting a differentiation with respect to the continuous arc length $s$ and with the condition that $|\uv(s)|=a$ for all $s$. For discrete chains the relationship between $\kappa$ and $l_p$ was identified as \cite{Muralidhar:2014fy} $l_p/a=[\kappa-1+\kappa\coth(\kappa)]/2[\kappa+1-\kappa\coth(\kappa)]$, with $l_p/a\approx\kappa-1/2+O(e^{-2\kappa})$ for $\kappa\to\infty$. The rigid inextensibility constraint makes analytical work with the KP model difficult in all but the simplest of cases, and a number of approximate methods have been applied to the model to extract experimentally relevant predictions from the model.

A particularly useful approximation is that of the Weakly Bending Rod (WBR) model\cite{Wagner:2007fc,MARKO:1995ka,Skolnick:1977jv}, which assumes the chain is sufficiently stiff that it can be approximated by small undulations about an axis.  Each bond is assumed to satisfy $\uv\approx (\uv_\perp,\sqrt{1-\uv_\perp^2})\approx (\uv_\perp,1-\uv_\perp^2/2)$ with $|\uv_\perp|\ll 1$, providing a parametrization that satisfies the inextensibility of the chain to second order.  This leads to an approximate Hamiltonian $H_{WBR}=\frac{l_p}{2}\int_0^L ds \dot{\uv}^2_\perp$, with the WBR approximation automatically enforcing the inextensibility constraints to second order.  This approximation has been usefully applied to polymers under an external tension\cite{MARKO:1995ka} and charged molecules\cite{Skolnick:1977jv}, and more recently to cylindrically confined WLCs.  The WBR model is particularly well suited to cylindrical confinement, where one might expect the dominant axis of the chain to align with the axis of the cylinder for strongly confined chains.  Confinement is imposed\cite{Wagner:2007fc} by coupling the WBR Hamiltonian to a harmonic potential of strength $\gamma$, with the approximate form $H_{WBR}^{cyl}=\frac{l_p}{2}\int_0^L ds \ddot{\rv}_\perp^2(s)+\frac{\gamma}{2}\int_0^L ds \rv_\perp^2(s)$ and where $\rv_\perp(s)=\int_0^s ds'\uv_\perp(s')$ is the transverse position of the monomer at arc length $s$ along the backbone.   The transverse bending correlation function for this model can be computed exactly\cite{Wagner:2007fc,wagner_2004} in the WBR limit, with 
\begin{eqnarray}
\langle\uv_\perp(s)\cdot\uv_\perp(s')\rangle=\frac{l_d}{l_p}e^{-|s-s'|/l_d}\bigg[\cos\bigg(\frac{|s-s'|}{l_d}\bigg)-\sin\bigg(\frac{|s-s'|}{l_d}\bigg)\bigg]\label{WBRconfCorr}
\end{eqnarray}
where $l_d$ is identified as a length scale proportional to the deflection length\cite{Odijk:1983jha,Chen:2013dl,Chen:2016gy}, $l_d\propto (l_p R^2)^{1/3}$ for $R$ the radius of confinement, and $\langle \cdots\rangle$ denotes an equilibrium average.  Note that this implies the scaling of the transverse fluctuations must satisfy $\langle\uv_\perp^2\rangle\sim (R/l_p)^{2/3}$, a fact relevant in the next section.  Note also that eq. \ref{WBRconfCorr} has extrema at $|s-s'|=n\pi l_d/2$ for integer $n$, implying that the correlation functions will necessarily have the lower bound $\langle \uv_\perp(s)\cdot\uv_\perp(s')\rangle \ge -l_d e^{-\pi/2}/l_p\propto (R/l_p)^{2/3}$.  In the WBR approximation, the rod is nearly extended and one readily finds $\langle\uv_{||}(s)\cdot\uv_{||}(s')\rangle\approx1-\langle\uv_\perp^2\rangle=1-\frac{l_d}{2l_p}$.   The mean squared end-to-end distance for this model is likewise readily computed through the integral $R_{ee}=\langle (\rv_L-\rv_0)^2\rangle=\int_0^L dsds' \langle \uv(s)\cdot \uv(s')\rangle$.

The WBR model was successfully applied to cylindrical confinement\cite{Wagner:2007fc}, for which the dominant axis for the polymer can be assumed to be the same as the confinement axis.  A similar calculation can be performed for a weakly bending rod in slit confinement, although we are not aware of this application of the WBR model to slits in precisely this manner in the literature.  For a WBR confined to a slit, fluctuations in the confined and one unconfined dimension are both assumed small.  If the confinement is assumed to be applied in the $z$ direction and the dominant axis of the chain is the $x$ axis, the Hamiltonian in the WBR limit can be written $H_{WBR}^{slit}\approx \frac{l_p}{2}\int_0^L ds[\ddot y^2(s)+\ddot z^2(s)]+\frac{\gamma}{2}\int_0^L ds z^2(s)$.  In this limit the confined and unconfined dimensions are decoupled, and a bending correlation function $\langle u_z(s)u_z(s')\rangle$ for slit confinement can be computed using the same techniques $\langle \uv_\perp(s)\cdot\uv_\perp(s')\rangle$ for cylindrical confinement, leading to the prediction of a damped oscillatory behavior for the correlation functions as in eq. \ref{WBRconfCorr} but with slightly different coefficients (e.g. for cylindrical confinement the WBR model predicts $l_d\propto \left({l_p}/{\gamma}\right)^{1/4}$).

\subsection{\label{MeanFieldDef.sec}The mean field Hamiltonian for confined WLC's}

The WBR approximation requires the polymer to be very stiff to accurately model either a chain weakly confined to a cylinder ($\gamma\to 0)$ or a slit-confined chain (where a strongly confined chain is effectively two-dimensional).  In order to better understand the statistics of a confined WLC confined to a slit without an assumption of a very stiff polymer, we adopt a mean field model to capture the rigid constraints on average.  This approach was successfully applied to the determination of correlation functions for WLCs confined to the interior of a sphere, where constraints are satisfied on average using a quadratic Hamiltonian.  The Hamiltonian is taken to be
\begin{eqnarray}
\beta H=\frac{l}{2}\int_0^L ds \dot{\uv}^2(s)+\lambda_{||}\int_0^L ds\uv_{||}^2(s)+\lambda_\perp\int_0^L ds \uv_\perp^2(s)+\frac{k}{R^2}\int_0^L ds\rv_\perp^2(s)\label{fullHam}
\end{eqnarray}
where $R$ is the dimension of the confinement:  the cylinder's radius for two confined dimensions (cylinder, $d_c=2$) or half of the slit separation for one confined dimension (slit, $d_c=1$).   The Gaussian form of the Hamiltonian permits separation of the confined and unconfined terms with $\beta H=\beta H_{||}+\beta H_{\perp}$.  Here, the bending energy is in terms of a mean field persistence length $l$, rather than the physical persistence length $l_p$ in the KP model, discussed further below.  The free parameters $\lambda_{||}$, $\lambda_\perp$ and $k$ are chosen to satisfy
\begin{eqnarray}
\frac{1}{L}\bigg\langle \int_0^L ds \uv^2\bigg\rangle=1\qquad\frac{1}{L}\bigg\langle\int_0^L ds \uv_\perp^2\bigg\rangle= \bar u\qquad\frac{1}{LR^2}\bigg\langle\int_0^L ds \rv_\perp^2\bigg\rangle=\bar r\label{constraint}
\end{eqnarray}
where $R^2\bar r$ and $\bar u$ denote the mean squared position and bending in the transverse direction.  The first relation in Eq. \ref{constraint} satisfies the constraint of inextensibility (since the average mean squared length of the chain is $L$).   The statistics of the chain in the transverse direction are satisfied by the last two relations in Eq. \ref{constraint}, constraining the mean squared bending (via the unknown $\bar u$) and  position (via the unknown $\bar r$) in the transverse direction.   Note that the variable $\bar u$ is an undetermined parameter here, but the scaling $ \bar u\sim (R/l_p)^{2/3}$ is expected due to the WBR theory\cite{Wagner:2007fc}.  This scaling can be confirmed on the mean field level without holding the transverse fluctuations fixed, although the MF theory does not accurately reproduce the scaling coefficient (data not shown). In this paper, we will extract $\hat{u}$ and $\hat{r}$ from simulations and compare the predictions of the MF and WBR models.

Excluding endpoint effects that do not affect the bulk behavior of the chain (discussed further in the SI), the free energy of the system can be determined by integration of the Hamiltonian in eq. \ref{fullHam} over the confined and unconfined spatial dimensions.  The $(3-d_c)$ unconfined dimensions contribute a free energy\cite{Ha:1995db} $\F_{unconf}=-\log\int{{\cal{D}}}(\uv_{||})e^{-\beta H_{||}}=(3-d_c)L\sqrt{\lambda_{||}/2l}$, while the $d_c$ confined dimensions contribute\cite{Morrison:2009gb} $\F_{conf}=-\log\int{{\cal{D}}}(\uv_{\perp})e^{-\beta H_{\perp}}=d_cL(\omega_++\omega_-)/2$ for  $\omega_\pm=\sqrt{\frac{\lambda_\perp}{l}\left(1\pm\sqrt{\frac{2kl}{\lambda_\perp^2}}\,\right)}$.  The total free energy is then written in terms of the Lagrange multipliers $\lambda_\perp$, $\lambda_{||}$, and $k$ as 
\begin{eqnarray}
\frac{\F_{tot}}{L}\sim \frac{d_c}{2}\bigg(\omega_++\omega_-\bigg)+\frac{3-d_c}{2}\sqrt{\frac{2\lambda_{||}}{l}}.
\end{eqnarray}
The constraints are imposed by requiring $\partial \F_{tot}/\partial \lambda_\perp=L \bar u$, $\partial \F_{tot}/\partial \lambda_{||}=L(1-\bar u)$, and $\partial \F_{tot}/\partial k=LR^2\bar r$.  
After some tedious algebra, differentiation of the free energy with respect to $\lambda_{||}$, $\lambda_\perp$ and $k$ readily give the solutions
\begin{eqnarray}
\omega_\pm=\frac{d_c}{4\bar ul}\bigg(1\pm \sqrt{1-\frac{16l^2}{d_c^2R^2}\frac{ \bar u^3}{\bar r} }\bigg)\qquad \lambda_{||}=\frac{(3-d_c)^2}{8l(1-\bar u)^2}\label{bulkRoots}
\end{eqnarray}
This leads to the scaling $\F_{tot}\sim L/(l_pR^2)$ in the limit as $R\to \infty$, recovering the expected scaling laws \cite{Odijk:1978cn}.  Note that eq. \ref{bulkRoots} can be written $\omega_\pm=l_d^{-1}\pm i\omega_d$ with $l_d=4 \bar ul/d_c$ and $\omega_d =l_d^{-1}\sqrt{16l^2\bar u^3/R^2\bar r d_c^2-1}$ \cite{Morrison:2009gb}.  Here, we have used the suggestive notation of Re($\omega_\pm)=l_d^{-1}$ to indicate its relationship to the deflection length that will be made below.

\subsection{\label{MeanFieldCorrelation.sec}Correlation functions on the mean field level}

The transverse and longitudinal bond correlation functions can be readily computed by writing $\langle \uv_\perp(0)\cdot\uv_\perp(L)\rangle=\partial/\partial\alpha\ \langle e^{\alpha \uv_\perp(0)\cdot\uv_\perp(L)}\rangle|_{\alpha=0}$ and $\langle \rv_\perp(0)\cdot\rv_\perp(L)\rangle=\partial/\partial\alpha\ \langle e^{\alpha \rv_\perp(0)\cdot\rv_\perp(L)}\rangle|_{\alpha=0}$.  These averages have been computed previously for the unconfined\cite{Ha:1995db} and confined\cite{Morrison:2009gb} dimensions.  It is tedious to compute the correlation function for cylindrical or slit confinement using the solutions in eq. \ref{bulkRoots}, but, following the results of \cite{Hsieh:2008fa}, the eventual result is
\begin{eqnarray}
\langle \uv_\perp(s)\cdot\uv_\perp(s')\rangle_{bulk}&=& \bar u e^{-|\Delta s|/l_d}\bigg(\cos(|\Delta s|\omega_d)-\frac{1}{l_d\omega_d}\sin(|\Delta s|\omega_d)\bigg)\label{confCorr}\\
\langle \rv_\perp(s)\cdot\rv_\perp(s')\rangle_{bulk}&=&\bar r R^2e^{-|\Delta s|/l_d}\bigg(\cos(|\Delta s|\omega_d)+\frac{1}{l_d\omega_d}\sin(|\Delta s|\omega_d)\bigg)\label{confCorrPos}
\end{eqnarray}
Here, we have ignored the excess endpoint fluctuations in the chain (discussed further in the SI) and used the notation $\langle\cdots\rangle_{bulk}$ to explicitly indicate that this is an average for points far from the endpoints of the chain.   Noting that $1+l_d^2\omega_d^2=u l_d^2/\bar r$ it is straight forward to show that eq. \ref{confCorrPos} implies that $\partial^2 \langle \rv_\perp(s)\cdot\rv_\perp(s')\rangle_{bulk}/\partial s\partial s'=\bar r R^2({1+l_d^2\omega_d^2})/{l_d^2}\times e^{-L/l_d}\left(\cos(L\omega_d)-\frac{1}{l_d\omega_d}\sin(L\omega_d)\right)=\langle \uv_\perp(s)\cdot\uv_\perp(s')\rangle_{bulk}$, as is required.

The functional form of Eq. \ref{confCorr} closely mirrors that of Eq. \ref{WBRconfCorr} derived under the WBR approximation\cite{Wagner:2007fc}, but with the emergence of a new length scale, $\omega_d$.  While the deflection length $l_d$ still captures the exponential decay of correlations along the backbone, this {\em{different}} length scale $\omega_d^{-1}$ captures the oscillations along the backbone.  Assuming that $u=\eta (R/l)^{2/3}$ for $R\ll l$ (justified by the WBR model that showed $\langle \uv_\perp^2\rangle=(l_pR^2)^{1/3}$), we find that $\omega_d\propto l_d^{-1}$ but with a constant of proportionality that is not unity.  Rather, we find $\lim_{R\to 0}\omega_d=l_d^{-1} \sqrt{16\eta^3/d_c^2\bar r-1}$ even for infinitely long or for strongly confined chains.  The qualitative features of the MF and WBR models are thus expected to agree, but quantitative disagreements between the models are expected as confinement is relaxed and the WBR model's inherent assumption that $l_p\gg R$ becomes inapplicable.

\subsection{Longitudinal correlation functions}

Computing $\langle \uv_{||}(0)\cdot\uv_{||}(L)\rangle=\partial/\partial\alpha\ \langle e^{\alpha \uv_{||}(0)\cdot\uv_{||}(L)}\rangle|_{\alpha=0}$ is also straightforward using the mean field Hamiltonian\cite{Morrison:2009gb}, from which one readily finds
\begin{eqnarray}
\langle \uv_{||}(0)\cdot\uv_{||}(L)\rangle_{bulk}= e^{-(3-d_c)L/2l(1-\bar u)}\label{longitudinalCorr}
\end{eqnarray}
where again endpoint effects have been neglected.  Note that the mean field approach predicts an exponential decay in the in the correlations on a length scale proportional to the bare persistence length (rather than the deflection length).  This functional form for the longitudinal correlation functions are known to be incorrect for chains under strong cylindrical confinement\cite{Wagner:2007fc} due to the huge energetic cost of chain backfolding\cite{Muralidhar:2016bf,Chen:2017fyb,Odijk:2006ce}.  The global persistence length of a cylindrically confined chain grows very rapidly as a function of $R^{-1}$, and the WBR model more accurately predicts the longitudinal correlation functions (with $\langle \uv_{||}(0)\cdot\uv_{||}(L)\rangle\approx$ const for large $L$).  The failure of the mean field model to recover the correct behavior in a limit where the WLC becomes effectively one dimensional has been noted previously\cite{Ha:1997bg} in a different context.   Despite this failure of the mean field approach to model the longitudinal correlations for cylindrically confined WLCs, we will show in the next section that the transverse correlation functions under cylindrical confinement are still well captured by the theory, as are the statistics of a slit-confined WLC in both transverse and longitudinal dimensions.

\section{\label{Results.sec}Results}

\subsection{\label{simulation.sec}Simulation Methodology}
We will demonstrate the efficacy of the mean field approach by comparing its correlation function predictions to that of simulated confined worm-like chains. Chains are generated via Metropolis MC simulations utilizing two MC moves: reptation and pivot. Reptation moves are attempted with a 10\% probability, in accordance with \cite{Bleha:2018jcp}, otherwise a pivot move is attempted within the bulk of the chain. A move is accepted if the altered monomer remains within the confinement volume and the Metropolis criterion is satisfied, $p_{accept}=\min(1,e^{-\beta \Delta E})$ with $\Delta E$ the change in energy after the trial move. This is continued until the chain is equilibrated (\(\approx 3\times 10^7\) MC steps for 100 monomers). Cylindrically confined chains favor a more elongated state entropically and therefore the initial conformation is chosen to be elongated in the longitudinal direction. The initial configurations are drawn by generating bonds uniformly on the surface of the unit sphere (for slit confinement) or unit hemisphere with $u_z>0$ (for cylinder confinement) and accepting them if the position of the new monomer does not violate the confinement constraint.  This process is iterated until an initial chain of desired length (N-1) is created within the confinement geometry at which point statistics are extracted from the generated chain.  4$\times 10^4$ relatively short chains (consisting of 100 monomers) and 4$\times 10^3$ longer chains (250 monomers) were generated, are generated in this way, within both cylindrical and slit confinement geometries, for bending rigidities $\kappa\in\{5, 10, 20, 40\}$, and 100 different radii $R$. The chosen radii are exponentially distributed such that the strong confinement regimes with small radii are sampled at a higher rate than the weakly confined large radii.  

\begin{figure}[htbp]
\begin{center}
\includegraphics[width=\textwidth]{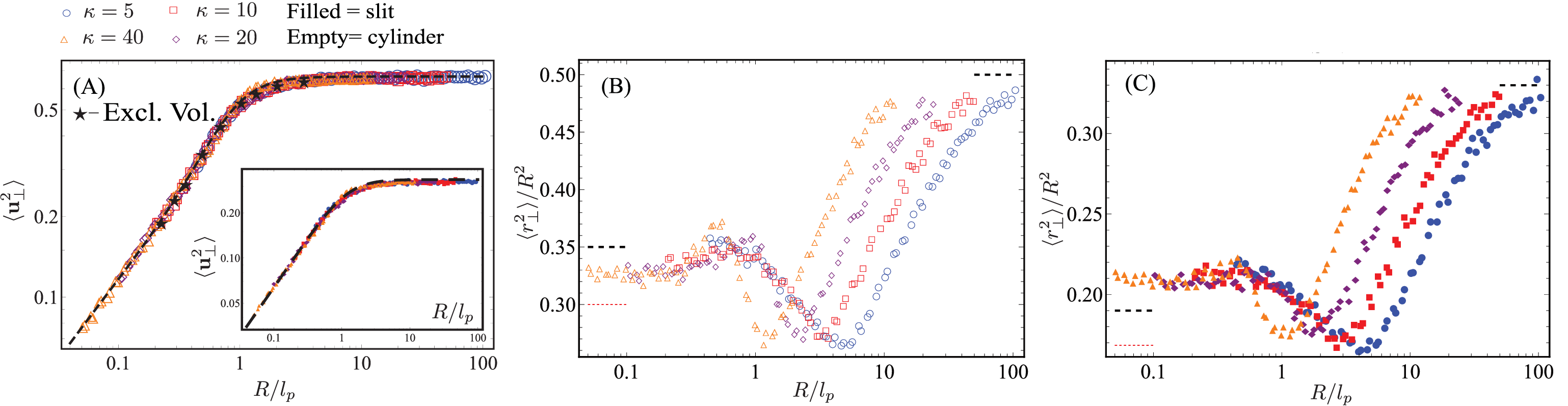}
\caption{(A) \(\bar{u}\) as a function of \(R/l_p\) for cylinder (main panel) and slit (inset) confinement. The excluded volume data was modified from \cite{Bleha:2018jcp}, and is in good agreement with our simulation results. The transition to the Odijk regimes begins around \(R/l_p\approx 1\) as expected. (B), (C) \(\bar{r}\) as a function of \(R/l_p\) for the cylinder and slit respectively. The dashed lines display the analytic limits for the confined rigid rod (red) and Gaussian chain (black). }
\label{uperpFig.fig}
\end{center}
\end{figure}

Our mean field approach requires two undetermined parameters if endpoint effects are ignored:  $\bar u=\langle \uv_\perp^2(N/2)\rangle$ (the transverse bending fluctuations far from the endpoints) and $\bar r=\langle \rv_\perp^2(N/2)\rangle/R^2$ (the transverse position fluctuations relative to the squared confinement radius far from the endpoints).  We are not aware of a simple analytical form for either value in the literature, although $\bar u$ may be determined for slit confinement by integrating a numerically determined distribution function\cite{Chen:2006cq}.  In this paper, we simply use the simulated values for $\bar u$ and $\bar r$ (determined from transverse position or bond vector of the $N/2$nd monomer) for varying $R$ and $l_p$.  These values are shown in Fig. \ref{uperpFig.fig}, and it is readily observed that $\bar u$ as a function of $R/l_p$ collapses onto a single curve for both small and large $R$ for all $\kappa$, suggesting a universal function may describe the transverse bending fluctuations for both cylindrical and slit confinement.  It is useful to determine an interpolation for $u$ as a function of $l_p/R$ and satisfying the known limits for small and large $R$: $u=1/3$ or 2/3 for slit and cylindrical confinement respectively when $R\to\infty$, and $u\sim (R/l_p)^{2/3}$ for small $R$.  Defining $x=R/l_p$ (with weak confinement in the limit of $x\to\infty$ and strong confinement when $x\to 0$), a family of simple functions that satisfies these limits is 
\begin{eqnarray}
\bar u(x)\approx\frac{d_c}{3}\bigg({1+\frac{1}{x^{2/3}g_{d_c}(x)}}\bigg)^{-1},\label{fittingUperpEq}
\end{eqnarray}
for some unknown $g_d(x)$ (with the assumption $g_d(0)\ne 0$).  If $x^{2/3}g_d(x)\to\infty$ as $x\to\infty$ (the weakly confined limit of large $R$), eq. \ref{fittingUperpEq} will yield the expected limits of $\bar{u}=d_c/3$ for weakly confined chains and $u\propto x^{2/3}$ for strongly confined chains.  For simplicity we take $g_d(x)\approx g^{(0)}_d+g_d^{(1)} x+\cdots$, and use Mathematica's NonlinearModelFit to determine the fitting coefficients from the simulated data, shown in Fig. \ref{uperpFig.fig}(a) for multiple values of $l_p$.  Fitting the combined data yields the best fit values
\begin{eqnarray}
\bar u_{slit}(l_p/R)\approx\frac{1}{3}\bigg(1+\frac{(l_p/R)^{2/3}}{0.943+1.453R/l_p+0.646(R/l_p)^2}\bigg)^{-1} \nonumber \\ \bar u_{cyl}(l_p/R)\approx\frac{2}{3}\bigg(1+\frac{(l_p/R)^{2/3}}{0.956+0.107R/l_p+3.089(R/l_p)^2}\bigg)^{-1}\label{bestfits}
\end{eqnarray}
These give the leading order term $u_{slit}\approx 0.31(R/l_p)^{2/3}$ and $u_{cylinder}\approx 0.64(R/l_p)^{2/3}$, in accordance with the expected scaling for small $R$. Eq. \ref{bestfits} is expected to be valid for sufficiently long (with $L/l_p\gg 1$) confined chains.  The fit produced by these equations is depicted as dashed lines in Fig. \ref{uperpFig.fig} and shows good agreement with the simulations for all sampled $R/l_p$ values. Additionally, the transverse statistics generated from excluded volume MC simulations (taken from \cite{Bleha:2018jcp}) match well to the results generated by our simulations by simply substituting $D_{exc.}-w=D_{ideal}$ where $D_{exc.}$ is the explicit diameter used in the excluded volume simulations, $w$ is the width of the beads in the excluded volume simulations, and $D_{ideal}=2R$ in our simulations. The agreement between our simulations and the excluded volume simulations is unsurprising as it mainly occurs over the Odijk regimes where the excluded volume effects become negligible in comparison to the effects of strong confinement.

We are not aware of an explicit prediction of the scaling of $\bar r=\langle \rv_\perp^2\rangle/R^2$ as a function of $R$ for a slit or cylindrically confined wormlike chain, although numerical studies of the shape of the transverse position distributions of confined WLCs have been performed for strongly confined chains\cite{Chen:2006cq,Burkhardt:1999fl}. We therefore choose to compare the simulation results for $\bar{r}$ to the expected values for a confined rigid rod and Gaussian chain as $R\rightarrow 0$ and $R\to\infty$ respectively, the details of which can be found in the SI. The dependence of $\bar r$ on $l_p/R$ is shown in Fig. \ref{uperpFig.fig} for cylindrical confinement, with a non-monotonic behavior as the confinement becomes stronger, and a bound of $0.26<\bar r<0.5$. For well-separated slits, the height of the rod will be uniformly distributed and $\bar r=R^{-3}\int_{-R}^R dzz^2=1/3$, consistent with the large $R$ behavior in Fig. \ref{uperpFig.fig}.  For a Gaussian chain confined between parallel plates\cite{Chen:2006cq}, the mean squared distance from the plates is shown in the SI to be $\langle \rv_\perp^2\rangle=\langle z^2\rangle=1-8/\pi^2\approx 0.19$, a value slightly larger than the minimum observed in Fig. \ref{uperpFig.fig} (of $\bar r_{slit}^{min}\approx0.17$). This suggests the minima in $\bar r$ correspond to the weakly confined limit of the chain (where confinement has an affect on its statistics but the Odijk scaling has not yet emerged).  In the limit of $R\to 0$, the rigid rod approximation yields $\bar{r}=1/6$.  A similar analysis shows that the limits for cylindrical confinement should be $\bar r=R^{-2}\int_0^r drr^3/\int_0^R drr=1/2$ for large $R$, and $\bar r\approx 0.35$ as $R\rightarrow 0$ for a Gaussian chain confined to a cylinder. The rigid rod approximation for $R\rightarrow 0$ has $\bar{r}=0.30$ (discussed further in the SI). In the cylindrical case the values for $\bar{r}(R/l_p)$ obtained from simulations lies within these limits for $R/l_p\rightarrow 0$ and approach the expected limiting value as $R\rightarrow\infty$ as show in fig. \ref{uperpFig.fig} (B).   

\subsection{Best fit correlation functions from the mean field and weakling bending rod predictions}

\begin{figure}[htbp]
\begin{center}
\includegraphics[width=\textwidth]{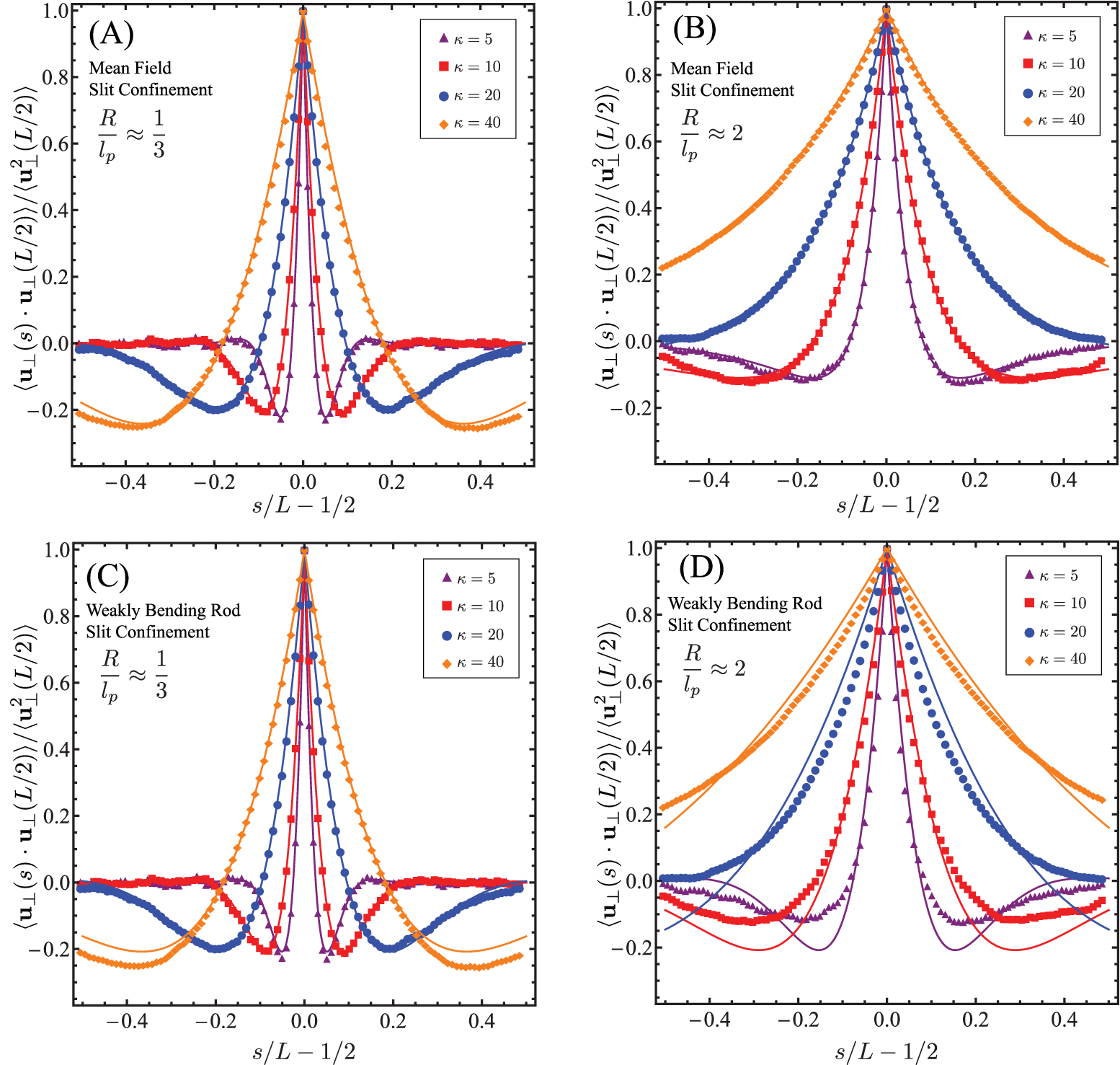}
\caption{Transverse bond correlations for slit confinement averaged over 40,000 configurations; (A),(B) Best fit of the normalized transverse correlations using equation 6 at \(R/l_p\approx1/3\) and \(R/l_p\approx2\) respectively. (C),(D) Best fit of the normalized transverse correlations using equation 1 at \(R/l_p\approx1/3\) and \(R/l_p\approx2\) respectively. Both models perform well under strong confinement (Odijk Regime) while the Mean field model proves more accurate under weaker confinement (beyond the backfolded Odijk regime). }
\label{SlitFit.fig}
\end{center}
\end{figure}

\begin{figure}[htbp]
\begin{center}
\includegraphics[width=\textwidth]{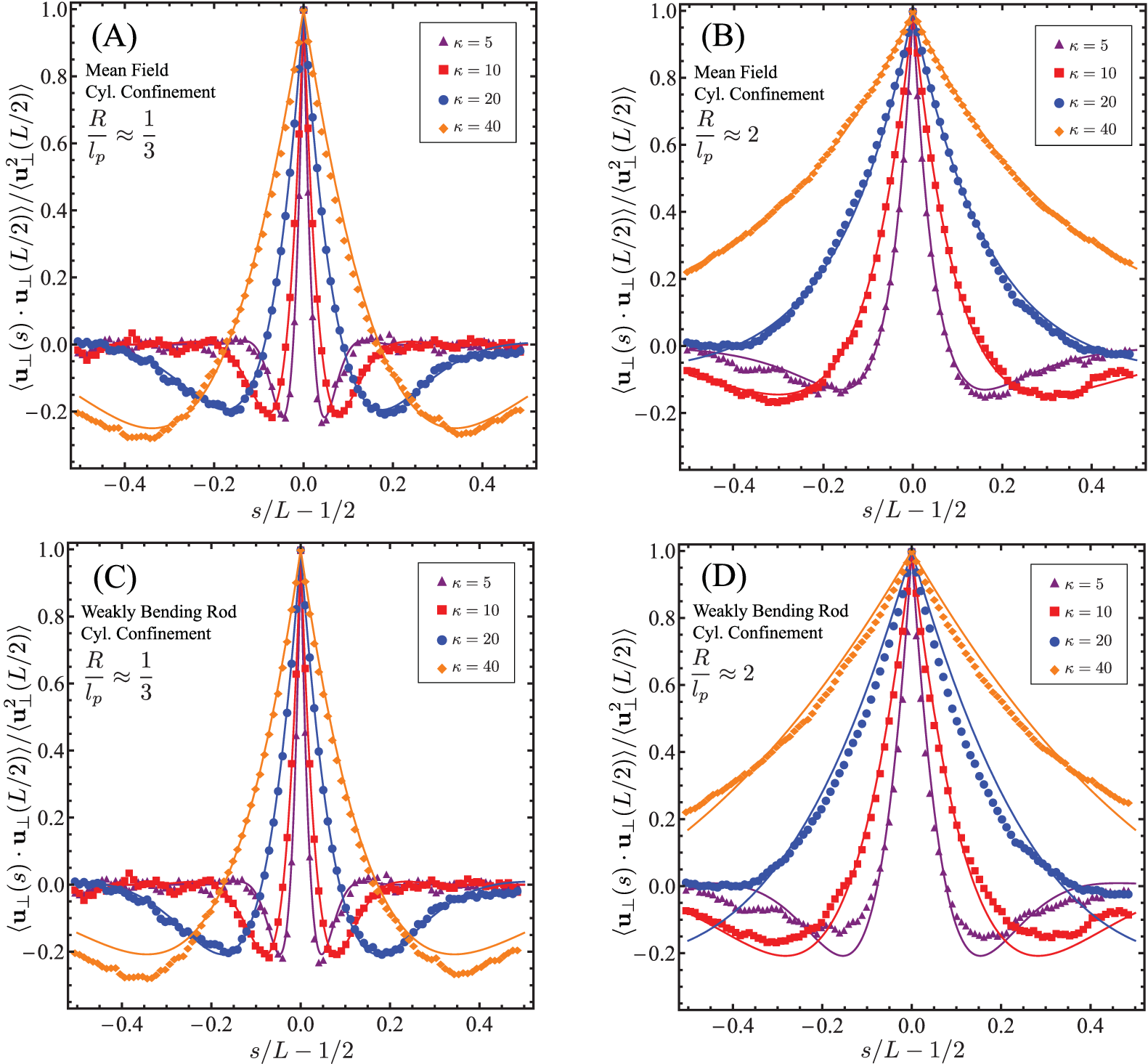}
\caption{As in fig. \ref{SlitFit.fig} the transverse bond correlations for a cylinder are shown along with the MF and WBR model predictions. Once again the MF and WBR model provide good fits within the Odijk regime yet the WBR model fails to provide an adequate fit beyond the backfolded Odijk regime highlighting the robust nature of the MF model in differing confinement geometries. }
\label{CylFit.fig}
\end{center}
\end{figure}

In order to assess the accuracy of the mean field approach, in Figs \ref{SlitFit.fig} and \ref{CylFit.fig} we show the best fit to the simulated transverse bending and position correlation functions for cylindrical and slit confinement using eqs. \ref{confCorr}.  For each value of $\kappa$ and $R$, the deflection length and frequency $l_d$ and $\omega_d$ are taken as two fitting parameters.  In both figures, the MF theory and simulations are in excellent quantitative agreement over the full range of $|s-s'|$ for strong confinement (with $R/l_p\approx 1/3$, depicted in Figs \ref{SlitFit.fig}(A) and \ref{CylFit.fig}(A)).   The WBR theory, which has a single fitting parameter ($l_d$) instead of two parameters of the MF approach, also accurately captures the qualitative behavior of the correlation functions for strong cylindrical and slit confinement (Figs \ref{SlitFit.fig}(C) and \ref{CylFit.fig}(C)). For intermediate confinement (with $R/l_p\approx 2$, depicted in \ref{SlitFit.fig}(B,D) and \ref{CylFit.fig}(B,D)), where the Odijk scaling of $\beta\Delta F\sim L/l_d$ is no longer expected to hold, the WBR approach no longer displays quantitative agreement with simulations than for strong confinement (although the best fit for the MF model remains accurate).  It is of course unsurprising that a fitting with two free parameters is able to produce a better overall fit of the data, but Figs. \ref{SlitFit.fig} and \ref{CylFit.fig} demonstrate that the additional length scale $\omega_d^{-1}$, capturing an oscillation length scale that is distinct from the exponential decay, is necessary to quantitatively describe the statistics of a confined wormlike chain.

\begin{figure}[htbp]
\begin{center}
\includegraphics[width=\textwidth]{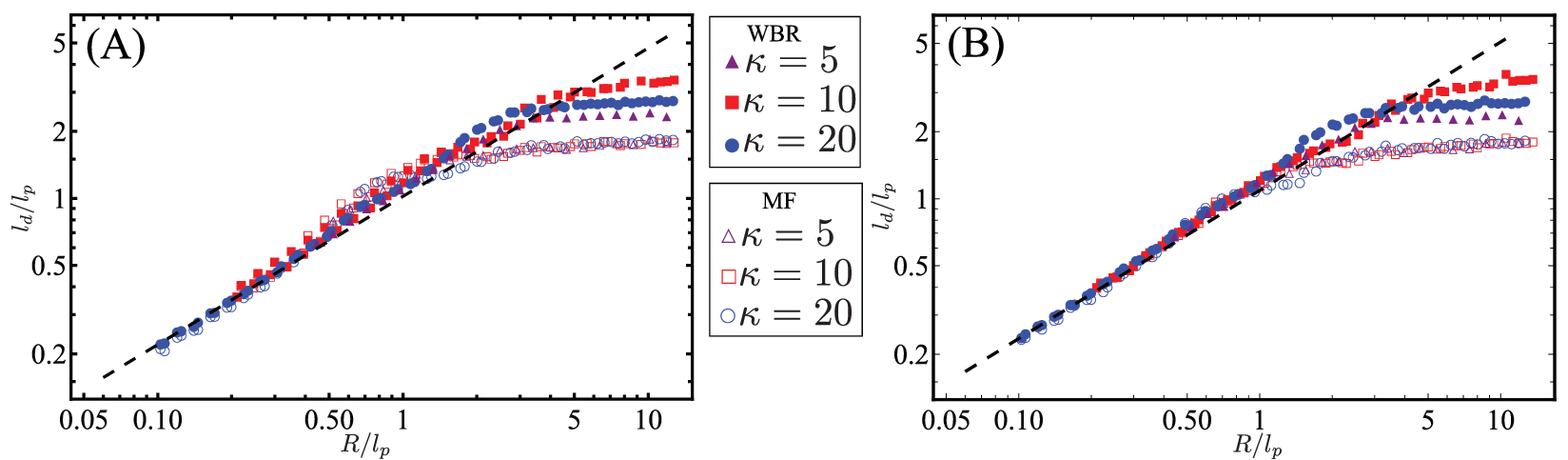}
\caption{Extracted values of the deflection length \(l_d\), from best fits, normalized by \(l_p\) as a function of \(R/l_p\) for (A) cylindrical and (B) slit confinement. The divergence of predictions of the two models for \(l_d\) occur at the end of the backfolded Odijk regime \(R/l_p\approx1\). As the confinement strength is decreased the weakly bending rod model predicts a deflection length which is dependent upon \(R/l_p\) while the Mean field model predicts a global deflection length independent of \(R/l_p\). The dashed lines are a guide to the eye for the expected scaling \(l_d/l_p\propto (R/l_p)^{2/3}\). The data has been truncated to focus on the transition near the backfolded Odijk regime. }
\label{Scaling.fig}
\end{center}
\end{figure}

The terms associated with the exponential decay in eqs \ref{WBRconfCorr} (for WBR) or \ref{confCorr} (for MF) are identified as the deflection length, and it is expected that $l_d\sim (l_pR^2)^{1/3}$ for strongly confined chains.  In Fig. \ref{Scaling.fig}, the behavior of the best fit values of $l_d$ as a function of the strength of confinement are shown for the slit and cylinder (in Fig. \ref{Scaling.fig}(A) and (B) respectively).   For strong slit and cylinder confinement, the measured values of $l_d$ do not differ significantly between the two theories and the scaling $l_d\propto (l_pR^2)^{1/3}$ is consistent across both theories.  For confnement beyond the Odijk regimes, there is a more significant difference between the theoretical approaches (with the MF approach generally giving a universal estimate for $l_d$, independent of $R/l_p$).  The solid lines show the scaling $l_d\sim l_p^{1/3}R^{2/3}$ as guides to the eye. It is clear that the MF approach predicts a global parameter $l_d$, while the WBR approach has an $l_d$ which is dependent upon $R/l_p$ beyond the Odijk regimes. 

\subsection{\label{transverse.sec}Effective longitudinal persistence length for varying confinement strength}

Longitudinal bending correlations are predicted to decay exponentially on a length scale proportional to the mean field persistence length $l$ in eq. \ref{longitudinalCorr}, with a constant of proportionality dependent on the transverse bending fluctuations $\bar u=\langle \uv_\perp^2\rangle$.   As discussed above, an exponential decay in the longitudinal bending correlations has been shown to be incorrect in the limit of $l/R\gg 1$ for cylindrical confinement (where $\langle \uv_\perp(s)\cdot\uv_\perp(s')\rangle\approx\mbox{const}$).   A cylindrically confined WLC behaves as an approximately one-dimensional system (having bonds pointed in the $\pm$z direction), with an initial exponential decay saturating to a constant value\cite{Wagner:2007fc}.  A continuum Hamiltonian as in eq. \ref{fullHam} cannot accurately model such a discrete system.  No such failure in the longitudinal correlations is expected for WLCs confined to slits, where the correlation functions converge on a three dimensional system for $R\to\infty$ (where $\langle\uv_{||}(s)\cdot\uv_{||}(s')\rangle=e^{-|s-s'|/l_p}$), and to a two dimensional system for $R\to 0$ (where $\langle\uv_{||}(s)\cdot\uv_{||}(s')\rangle=e^{-|s-s'|/2l_p}$).

\begin{figure}[htbp]
\begin{center}
\includegraphics[width=\textwidth]{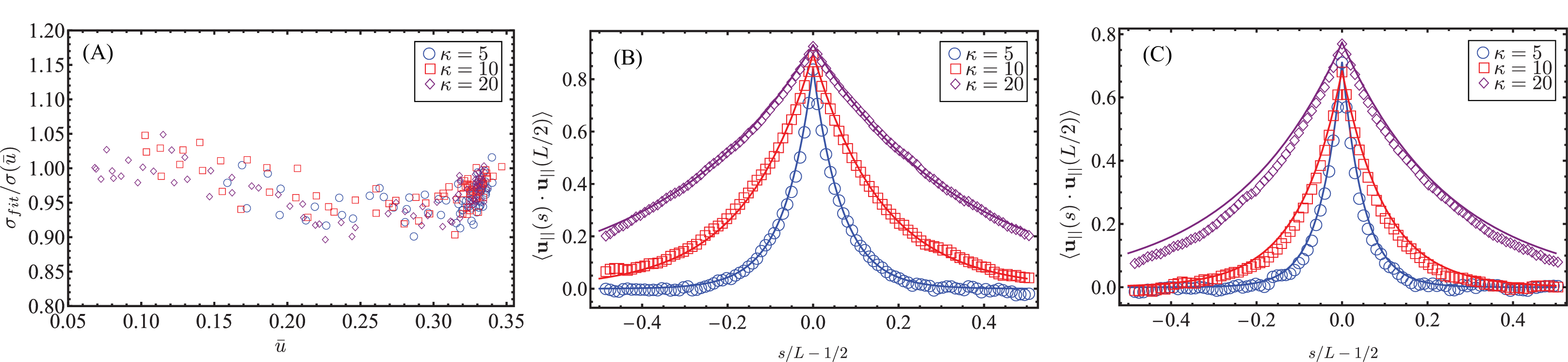}
\caption{(A) Comparisons of the values of $\sigma$ gained from the linear interpolation described in the text and by best fit of the simulation data. (B),(C) Show the fits of the longitudinal bond correlations gained from using the linear interpolation of \(\sigma\) at \(\bar{u}\approx0.15\) (B) and \(\bar{u}\approx0.23\) (C) where the deviation between the fit and linear interpolation is the greatest (\(\approx10\%\)). }
\label{mflp.fig}
\end{center}
\end{figure}

\begin{figure}[htbp]
	\begin{center}
		\includegraphics[width=.8 \textwidth]{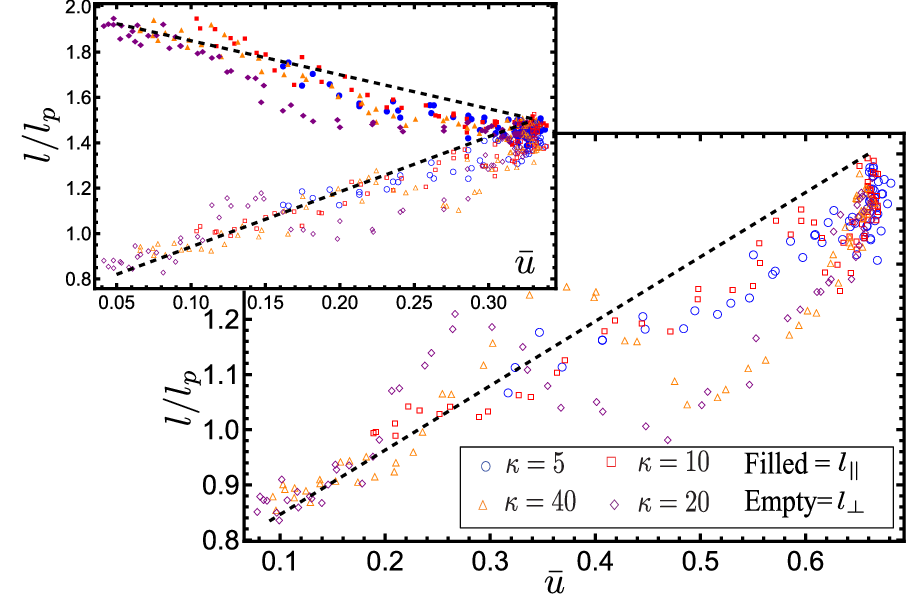}
		\caption{The mean field persistence length $l$ as a function of $\bar{u}$ extracted by fitting the transverse (filled symbols) or longitudinal (open symbols) for the cylinder- (main panel) and slit- (inset) confined systems.   The linear interpolations described in the text are represented by the black dashed lines.}
		\label{lmfvubar.fig}
	\end{center}
\end{figure}

It is important to note that the exponent in eq. \ref{longitudinalCorr}, $\langle \uv_{||}(s)\cdot\uv_{||}(s')\rangle=e^{-|s-s'|/l(1-\bar u)}$  {\em{does not}} satisfy the expected behavior in both of these limits for slit confined chains either.  For $R\to 0$ (where  $\bar u\to 0$), the correlation function converges to $\langle \uv_{||}(s)\uv_{||}(s')\rangle=e^{-|s-s'|/l}$ and for $R\to\infty$, where $\bar u\to \frac{1}{3}$, the correlation function converges to $e^{-3L/2l}$.  The failure to precisely recover the known decay length is a previously recognized issue with this mean field approach\cite{Morrison:2009gb}, and is resolved by modifying the mean field persistence length, $l$, so that the correlation function satisfies the expected limits.  
The mean field persistence length will satisfy $l=l_p\times f(\bar u)$ for some unknown function $f(\bar u)$, with $f(0)=2$ (matching the two dimensional decay length of $l=2l_p$) and $f(1/3)=3/2$ (matching the three dimensional decay length of $l=3l_p/2$).  A linear interpolation that satisfies the boundary conditions is $l_{||}(\bar u)=2l_p(1-\frac{3}{4}\bar u)$, with the decay of the longitudinal correlations thus expected to satisfy 
\begin{eqnarray}
\langle \uv_{||}(s)\cdot\uv_{||}(s')\rangle\propto e^{-|\Delta s|/l_{||}(\bar u)(1-\bar u)}=e^{-|\Delta s|/2l_p(1-3\bar u/4)(1-\bar u)}\equiv e^{-|\Delta s|/\sigma(\bar u)}.\label{sigmadef}
\end{eqnarray}
Fig. \ref{mflp.fig}(A) shows the best fit correlation length, $\sigma_{fit}$, for all simulated values of $\kappa$ and $R$, normalized by the expected $\sigma(\bar u)$.  The deviation between the fit and predicted length scales is $\lesssim 10\%$  over the observed values of $\bar u$.  While a more detailed model may be able to better fit the simulations by utilizing a free parameter for fitting (e.g. fitting $l_{||}/l_p$ using a third-order polynomial gives an improvement to $\lesssim 5\%$ deviation; data not shown), the simple approximation implemented here solely uses the predictions in eq. \ref{longitudinalCorr} combined with an interpolation dictated by the expected boundary conditions.  The quality of the agreement between the exponential fit $e^{-|s-s'|/\sigma(\bar u)}$ are shown in Fig \ref{mflp.fig}(B) for strongly confined examples (with $R=2a$) and (C) for values of $R$ for which $\bar u=\langle u_\perp^2\rangle\approx 0.25$ (where the deviation in Fig. \ref{mflp.fig}(A) are greatest).  In both cases, the theoretical predictions are in good agreement with the simulations.  We likewise find the linear approximation for $l_{||}/l_p$ agrees fairly well with the simulated values, shown in the filled symbols of the inset of Fig. \ref{lmfvubar.fig}.  The mean field longitudinal decay length of $\sigma(\bar u)$ and equivalently the mean field persistence length $l_{||}$ adequately describes the longitudinal correlation length of slit-confined wormlike chains while requiring only one parameter:  the mean transverse bending $\bar u=\langle u_\perp^2\rangle$.

In the previous paragraph, we used the notation $l_{||}=2l_p(1-\frac{3}{4}\bar u)$ to indicate the mean field persistence length as determined by fitting the bond correlation function under slit confinement in the unconfined dimensions.  One could also determine the mean field persistence length using eq. \ref{confCorr}, applicable to both cylindrical and slit confinement, with the best fit value denoted $l_\perp$ using the transverse correlation function.  The transverse mean field persistence length $l_{\perp}$ will not be equivalent to the longitudinal one $l_{\parallel}$ for highly confined systems.  This is immediately evident for cylindrical confinement:  $l_{||}$ is not well defined (as the exponential decay is known to be incorrect) and $l_{\perp}=2l_p/3$ is required in the limit of $R\to\infty$ to recover the expected WLC decay length.  For slit confinement, we expect that $l_{||}=l_\perp=1/3$ in the limit of $R\to\infty$, but have no a priori reason to predict that the mean field persistence lengths must agree for $R\to 0$.  To determine $l_\perp$, we perform a best fit on both the cylindrical and slit confined simulations in the transverse directions (using eq. \ref{confCorr}), from which we determine $l_d=4\bar ul_\perp/d_c$ (as described in Sec. 2.2).  In Fig. \ref{lmfvubar.fig}, we find $l_\perp/l_p$ varies in an approximately linear fashion the open symbols in the main panel and inset), but with different coefficients from $l_{||}$.  We find that the best linear interpolation for $l_\perp$ from our simulation data is $l_\perp^{slit}\approx l_p(0.7+2.4\bar{u})$ for the slit and $l_\perp^{cyl}\approx l_p(0.7+\frac{3}{2}\bar{u})$ for the cylinder.  

\subsection{\label{Infer.sec}Inferring transverse fluctuations from the longitudinal end-to-end distance in slit-confined chains}

It is often the case experimentally that confined polymers are readily visible in some dimensions, but difficult to directly observe in others.  Generally, a direct observation of chain statistics in the $x-y$ plane may be possible for either cylindrical (or channels with a square cross-section) or slit confinement\cite{Werner:2018dm,Yeh:2016bs}. Observation in a confined dimension $z$ may be infeasible, since fluctuations in the chain are on the confinement length scale $R$ and not resolvable using the techniques employed in a typical fluorescence experiment.  For cylindrically confined chains, one expects the fluctuations along the directly-observable $y$ axis are the same as along the unobservable $z$ axis by symmetry.  However, slit-confined chains have no such symmetry, and an observation in the $x-y$ plane gives no immediate information about transverse fluctuations.  The functional form of eq. \ref{sigmadef} shows the longitudinal correlation function {\em{does}} have a dependence on $\bar u$, which is the mean squared bending in the transverse direction.  Thus, by extracting the parameter $\bar u$ experimentally from a longitudinal observation, the transverse statistics can be inferred for strong confinement.  

\begin{figure}[htbp]
\begin{center}
\includegraphics[width=\textwidth]{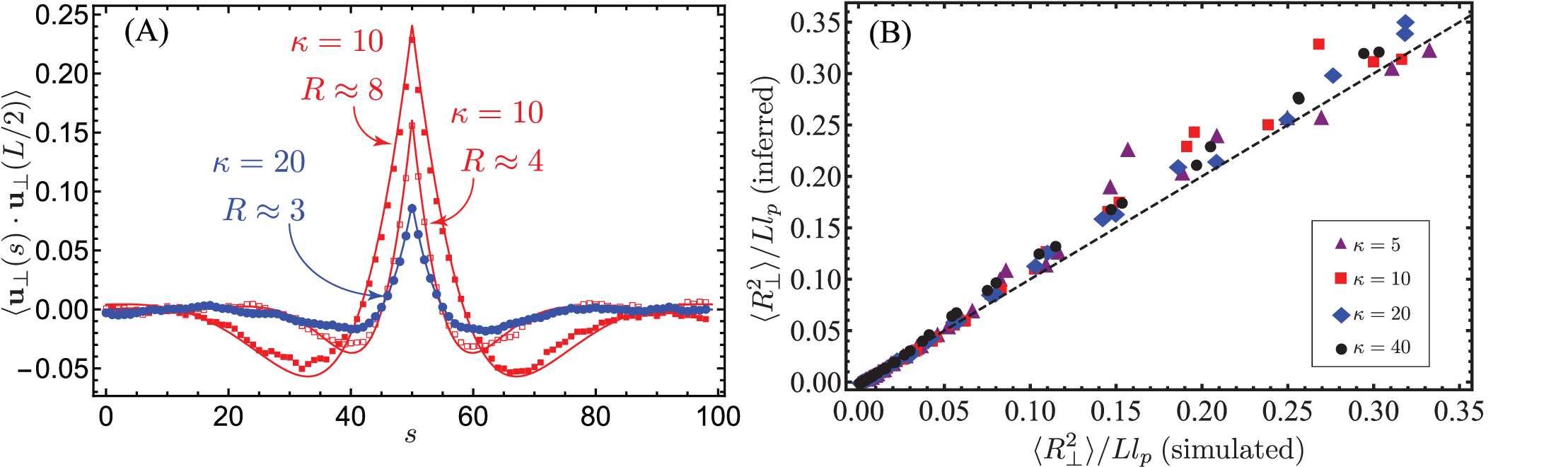}
\caption{(A) Comparison of the analytic expectations and simulated transverse bond correlations. The values for substitution into equation 6 were obtained by extracting \(\bar{u}\) as described in the text. (B) Comparison of \(\langle R_{\perp}^2 \rangle/Ll_p\) for various \(\kappa\) obtained from simulations by inferring $\bar u$ from the observed \(\langle R_{\perp}^2 \rangle\) as described in the text. }
\label{InferredRee.fig}
\end{center}
\end{figure}

For a WLC confined in a slit, the parameter $\bar u$ can be directly measured through fitting of a correlation function (in eq. \ref{sigmadef}) or more easily by measuring the longitudinal mean-squared end-to-end distance,
\begin{eqnarray}
\frac{\langle\Rv_{||}^2\rangle}{L^2}=\int_0^L \langle \uv_{||}(s)\cdot\uv_{||}(s')\rangle=2(1-\bar u)\frac{\sigma (\bar u)}{L}\bigg(1-\frac{\sigma(\bar u)}{L}\left[1-e^{-L/\sigma(\bar u)}\right]\bigg)\label{longRee}
\end{eqnarray}
computed from eq. \ref{sigmadef}.  $\langle \Rv_{||}^2\rangle$ is an observable requiring only a measurement of the separation between endpoints observed in the $x-y$ plane, without having to directly observe the confined dimension.  An experimental measurement of $\langle \Rv_{||}^2\rangle$ can thus be used to determine $\frac{\sigma(\bar u)}{L}$ from eq. \ref{longRee}, from which the purely transverse quantity $\bar u=\langle \uv_\perp^2\rangle$ can be determined (assuming $L$ and $R$ are known {\em{a priori}}).  The statistics of the chain in the transverse direction depend on the two unknown parameters $\bar u$ (which can be inferred using eq. \ref{longRee}) and as well as the unknown $\bar r$.  While this latter quantity cannot be estimated from the longitudinal statistics, for strong slit confinement we expect $\bar r\approx 0.2$.  It is thus possible to estimate the transverse correlation function using the inferred value of $\bar u$ and the interpolation for $l_\perp(\bar u)$ described in the previous section, as shown in Fig. \ref{InferredRee.fig}(A). The agreement between the simulated and inferred statistics are fair over a wide range of $l_p$ and $R$.   We also see that the inferred transverse end-to-end distance $\langle R_\perp^2\rangle=\int_0^L dsds'\langle u_\perp(s)u_\perp(s')\rangle$ (shown in Fig. \ref{InferredRee.fig}(B)) is well predicted by inferring $\bar u$ solely from the the longitudinal statistics.   These data suggest that the MF theory can be used to accurately describe the fluctuations of the chain in the confined dimension of the slit, even if they cannot be directly observed experimentally.

\section{Conclusions and Discussion}

In this paper, we have used a mean field theory to analytically determine the transverse and longitudinal position and bending correlation functions for wormlike chains confined to cylinders and slits.  The model predicts the emergence of two length scales in the correlation function:  an exponential decay length associated with the well-known deflection length ($l_d$), and a distinct length scale associated with oscillations in the correlation functions ($\omega_d^{-1}$) previously unreported in the literature.  The predicted correlation functions are in excellent agreement with the results of MC simulations, confirming the accuracy of the model.  The mean field model is also robust enough to effectively predict the correlation functions across the Odijk \emph{and} backfolded Odijk regime where the WBR model becomes less accurate.

In this paper, we have predicted the frequencies $\omega_\pm=l_d^{-1}\pm i\omega_d$, with the new length scale $\omega_d^{-1}$ not previously noted in the literature.  Because the free energy $\F\sim L(\omega_1+\omega_2)\sim L/l_d$ (with the $\omega_d$ terms cancelling), our theory is necessarily consistent with the already well-developed work in the field of cylinder- or slit-confined wormlike chains.  The emergence of the new length scale $\omega_d^{-1}$ thus does not indicate an inconsistency between the current work and the existing literature:  extensive contributions to the free energy are dominated by the deflection length.  The predictions made here are thus suggestive that {\em{local}} correlations and fluctuations may involve this additional length scale for intermediate values of $R$.  This result may have important implications for nanopore sequencing techniques, for which noise statistics may depend strongly on the local fluctuations in the chain.  The general agreement between the mean field and WBR theories for strong confinement in Fig. \ref{Scaling.fig} as well as the reasonable agreement between the fits in Fig. \ref{SlitFit.fig} suggest $\omega_d^{-1}\approx l_d$ for a WLC confined between plates, and largely validates both theoretical approaches. However, the MF method is capable of effectively predicting the behavior of semi-flexible polymers across multiple confinement regimes. It's relative simplicity and robust nature making it a natural choice for treatments of biomolecules confined within or beyond the Odijk and backfolded Odijk regimes. 

Despite the useful predictions made for transverse correlations in cylindrical confinement, this model cannot capture the expected statistics of the linear extension of a cylindrically confined wormlike chain (which has the predicted scaling of $\langle |Z|\rangle/L-1\sim l_d/l_p$).  This failure is due to the theory's incorrect prediction of an exponentially decaying longitudinal correlation length rather than the expected $\langle \uv_{||}(s)\uv_{||}(s')\rangle$=const for cylindrical confinement.  This issue arises because the mean field theory assumes a probability distribution of $u_{||}(s)$ that is peaked at 0, rather than near $\pm 1$ in the limit of $R\to0$.  Overcoming this problem would require additional assumptions in the theory to bias the most probable value of $u_{||}(s)$ towards a nonzero value, which the assumptions underlying the WBR theory or quasi-one dimensional theories\cite{Huang:2015hf} might make possible.   A hybrid theory, implementing the mean field approach with additional assumptions from these other theories, would also be useful in the context of backfolding of confined wormlike chains\cite{Muralidhar:2014bla,Odijk:2006ce}, where strongly confined chains sometimes experience tight bends (changing their direction in the channel) over a length scale termed the `global' persistence length, $g$.  Investigation into the determination of the global persistence length as well as three-dimensional confinement is planned for future work.

The statistics for slit-confined chains are accurately predicted for both the transverse and longitudinal dimensions (unlike cylindrical confinement), and in this paper we have suggested a method by which predictions about the transverse statistics of the chains can be made using only direct observations of the longitudinal dimensions.  This theory can in principle be extended to account for external tensions or fields, which have previously been realized experimentally\cite{Yeh:2016bs}, to better understand the the stretching of confined DNA in single molecule experiments.  Throughout this paper we have relied on an assumption that electrostatic and excluded volume effects are perturbative, with the underlying approximation that the  contribution of those interactions to the persistence length ($l_{total}=l_{p}+l_{int}$) is independent of $R$.  This theoretical approach\cite{Ha:1995iy} can be applied to determine $l_{el}(R)$ for slit-confined charged WLCs to validate this underlying assumption of a constant $l_{el}$, a possible avenue for future work.   It is worth emphasizing that the method of inferring $\bar u$ from observable statistics in the to predict $\langle R_\perp^2\rangle$ does not depend on the assumption that $l_p$ is known, but rather under the assumption that the wormlike chain model can be used at all.  Assuming the statistics of an interacting, stiff chain can be recovered using a renormalized wormlike chain, we expect the MF theory and inference method described here to be useful.  


\section{SI}

\subsection{Endpoint effects in the simulations for large $\kappa$}

\begin{figure}[htbp]
\begin{center}
\includegraphics[width=\textwidth]{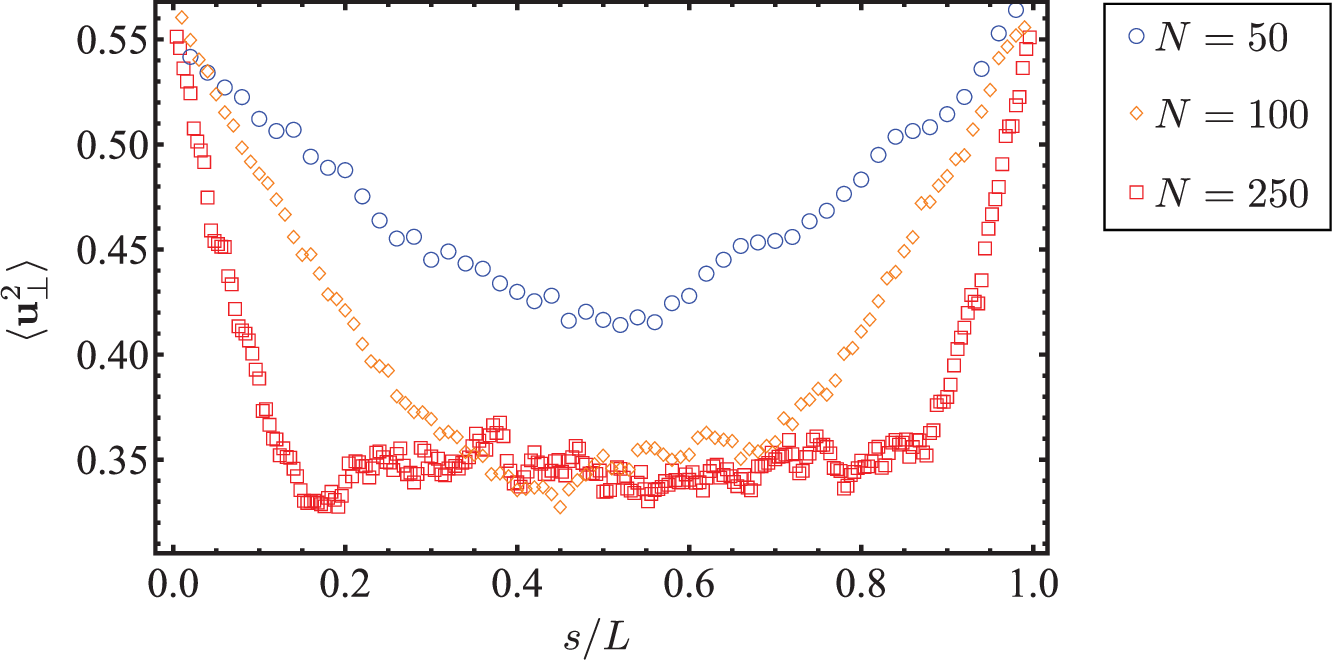}
\caption{Illustration of endpoint effects for \(\kappa=40\), \(R=20\) for \(N=50\), \(N=100\), and \(N=250\). The pervasiveness of the effects on \(\langle \bold{u}_{\perp}^2\rangle\) is \(\approx l_p\).}
\label{SiCorrFig.fig}
\end{center}
\end{figure}

\begin{figure}[htbp]
	\begin{center}
		\includegraphics[width=\textwidth]{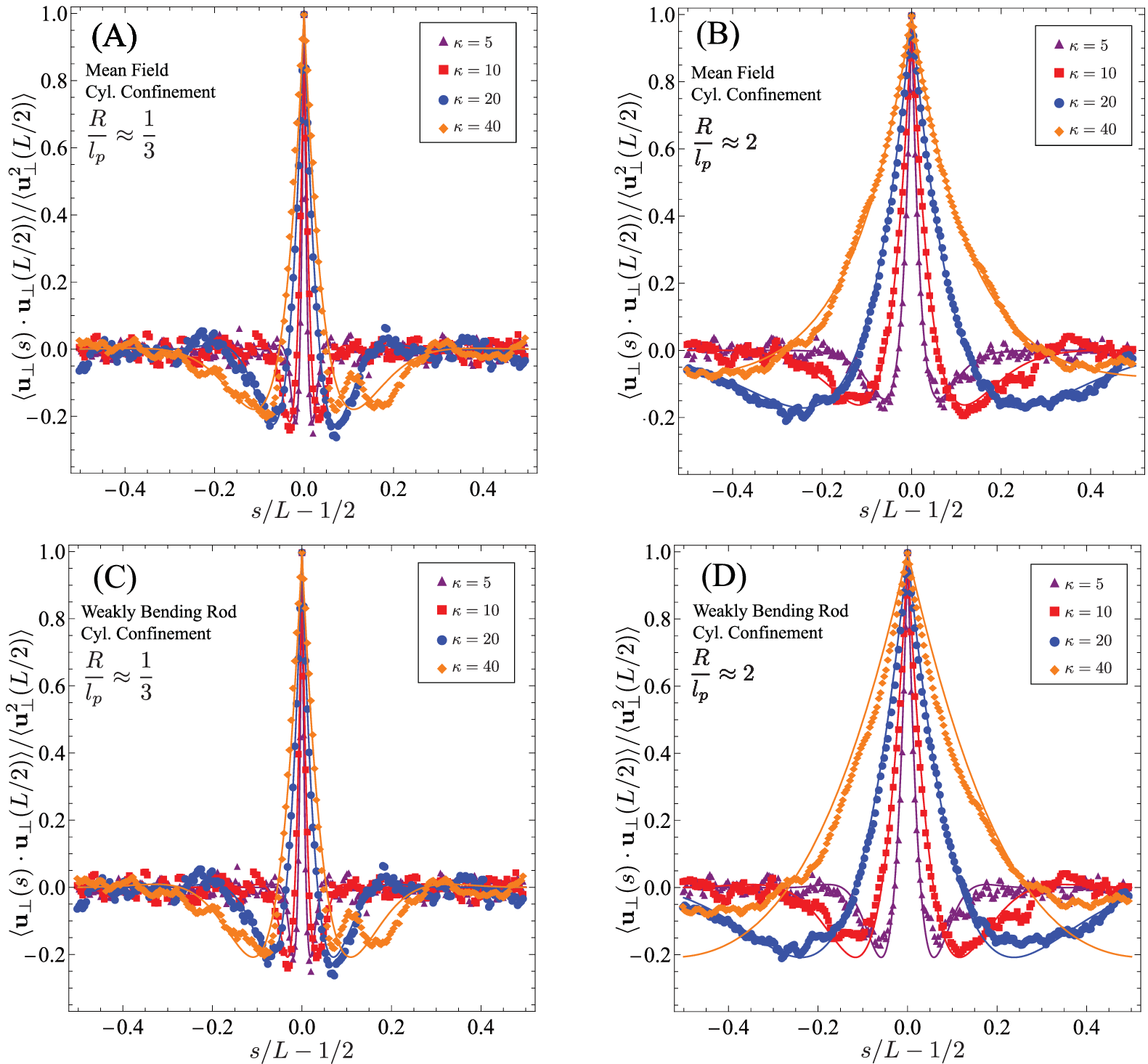}
		\caption{For N=250 the MF theory remains robust across confinement regimes and all values of $\kappa$.}
		\label{CylFit250.fig}
	\end{center}
\end{figure}

\begin{figure}[htbp]
	\begin{center}
		\includegraphics[width=\textwidth]{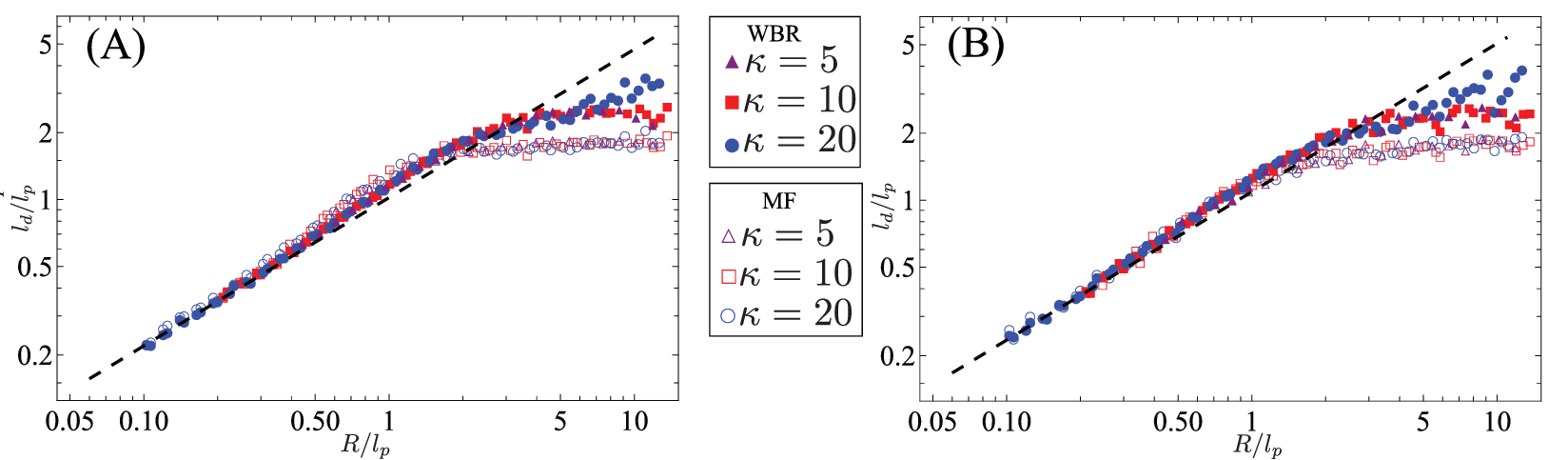}
		\caption{Predicted deflection length comparisons for N=250; the MF theory still predicts a length independent of $R/l_p$ beyond the Odijk regime while the WBR does not.}
		\label{deflecscaling250.fig}
	\end{center}
\end{figure}

For slit and cylindrical confinement, the mean transverse fluctuations $\bar u=\langle\uv_\perp^2\rangle$ collapse onto a single curve for all $\kappa=5$.  The predictions of the MF model for $\kappa\ge30$ become less accurate due to endpoint effects.  This is due to the finite length of the chains, as shown in SI Fig \ref{SiCorrFig.fig}.  The mean field model assumes the bulk behavior of the chain $\langle \uv_\perp^2\rangle\approx$const for points far from the endpoints of the chain, but SI Fig \ref{SiCorrFig.fig} shows that $\langle \uv^2(s)\rangle$ does not reach a saturating value for interior points of a cylindrically confined chain for $\kappa=40$ and $R/l_p\approx 0.5$.  These boundary effects prevent an accurate identification of $\bar{u}$ for strongly confined cylindrical chains.  Similar effects are observed for slit confined chains, shown in SI Fig. \ref{SiCorrFig.fig}(B).  Endpoint effects persist over a range on the order of $l_p$, and strongly confined stiff chains are expected to have bulk behavior as long as $L\gg l_p\gg R$.  The choice of $L=100a$ in these simulations makes the underlying assumptions of the MF model only approximate and a meaningful identification of $\bar{u}$ difficult by simply sampling the interior points.  It is worth emphasizing that the mean field parameters $l_d$ and $\omega_d$ can still be accurately used as fitting parameters (as in Fig. \ref{CylFit.fig}(A)) without requiring an estimate of $\bar u$. The endpoint effects do not affect the comparisons of the MF theory to the WBR (e.g. fig. \ref{CylFit.fig}) as shown in SI fig. \ref{CylFit250.fig}. Likewise the results shown in \ref{Scaling.fig} remain true for N=250 as well (SI fig. \ref{deflecscaling250.fig}).

\begin{figure}[htbp]
	\begin{center}
		\includegraphics[width=\textwidth]{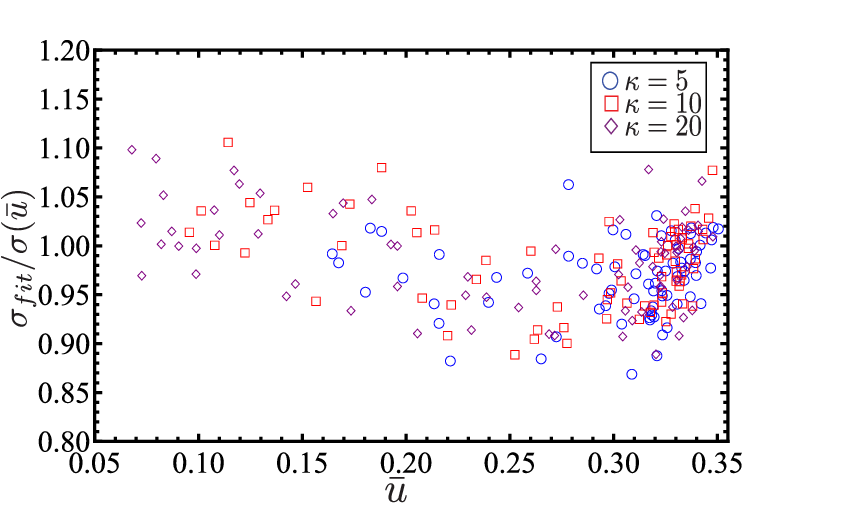}
		\caption{Comparison of values of $\sigma(\bar{u})$ fitted from simulation data and predicted values for $\sigma(\bar{u})$ for N=250. The larger deviations as compared to fig. \ref{mflp.fig} in the main text can most likely be attributed to the smaller data set produced for N=250.}
		\label{sigmau250.fig}
	\end{center}
\end{figure}

\subsection{Endpoint effects on the Mean Field approach}
The global constraints in eq. \ref{constraint} are generally insufficient to determine the statistics of a polymer on the mean field level, due to excess fluctuations at the endpoints\cite{Morrison:2009gb,Ha:1995db}.  These excess fluctuations are suppressed by imposing additional constraints that ensure $\frac{1}{2}\langle\uv^2(0)+\uv^2(L)\rangle=1$, $\frac{1}{2}\langle\uv_\perp^2(0)+\uv_\perp^2(L)\rangle=\bar u_e$, $\frac{1}{2R^2}\langle \rv_\perp^2(0)+\rv_\perp^2(L)\rangle=\bar r_e$, and $\langle\rv_\perp(0)\cdot\uv_\perp(0)\rangle=\langle\rv_\perp(L)\cdot\uv_\perp(L)\rangle$.  The first constraint ensures the endpoint bond vectors to have fixed length, the second and third constraints fix the mean squared transverse endpoint bond vectors and positions at $\bar u_e$ and $ \bar r R^2$ respectively, and the third arises from a symmetry argument\cite{Morrison:2009gb}.   Defining $C(\rv_0,\rv_L)=\delta(\uv_0^2+\uv_L^2-2)+\delta_\perp(\uv_{\perp,0}^2+\uv_{\perp,L}^2-2 u_e)+\gamma_1(\rv_{\perp,0}^2+\rv_{\perp,L}^2-2R^2 \bar r_e)+\gamma_2(\rv_{\perp,0}\cdot\uv_{\perp,0}-\rv_{\perp,0}\cdot\uv_{\perp,0})$ to account for the endpoint constraints with the six Lagrange multipliers $\lambda$, $\lambda_{||}$, $k$, $\delta$, $\gamma_1$ and $\gamma_2$, one can compute the partition function as\cite{Morrison:2009gb}
\begin{eqnarray}
Z=\int d\rv_0d\uv_0d\uv_Ld\rv_Le^{-C(\rv_0,\rv_L)}\int_{\shortstack{$\scriptstyle \rv(0)=\rv_0,\rv(L)=\rv_L$\\$\scriptstyle \uv(0)=\uv_0,\uv(L)=\uv_L$}}\D[\rv(s)]e^{-\beta H[\uv(s)]-kL\bar r-\lambda L-\lambda_{||} L(1-\bar u)}\label{PathIntegral}
\end{eqnarray}
All constraints are satisfied on average by minimizing the free energy $\F=-\log(Z)$ with respect to each of the Lagrange multipliers.

The free energy is readily evaluated with these endpoint effects using an explicit evaluation of the partition function\cite{Morrison:2009gb}, leading to a bulky expression that is tedious to work with.  However, it is straightforward to see that the extensive contribution to the free energy unchanged by the inclusion of the endpoint effects in the limit $L\to\infty$.  This means that the mean field solutions for the bulk parameters $\lambda$, $\lambda_{||}$, and $k$ are unchanged for long chains.   Differentiation with respect to the endpoint Lagrange multipliers yields four equations that are also exactly solvable in the limit of $L\to\infty$, and one can show $\gamma_1=d_c(2\bar r_e^{-1}-\bar r^{-1})/4$, $\gamma_2=l \bar u/2R\bar r$,  $\delta=(3-d_c)(1-2\bar u+\bar u_e)/4(1-\bar u)(1-\bar u_e)$, and $\delta_{\perp}=(3 \bar u-d_c)/ \bar u(1- \bar u)\ -\ 2(3 \bar u_e-d_c)/\bar u_e(1- \bar u_e)$.  These mean field solutions allow a computation of many mean quantities of interest in terms of the phenomenological parameters $ \bar u$, $\bar r$, $ \bar u_e$, and $\bar r_e$.    In this paper we focus solely on bulk statistics (where endpoint effects can be neglected), since experiments typically focus on long chains and inclusion of these effects require the use of more free parameters than the two ($\bar u$ and $\bar r$) in the main text.  

\subsection{Confined Rod}

A rigid rod in strong cylindrical confinement is path limited due to the confining walls such that
\begin{eqnarray}
	\bar r=\frac{\int_{0}^{R}r^3\arcsin(2(R-r)/L)dr}{R^2\int_{0}^{R}r\arcsin(2(R-r)/L)dr}
\end{eqnarray}
which is valid for $L>2R$. If the confinement is weaker such that $L\le 2R$ the rod can now rotate freely within the cylinder and is only constrained near the walls i.e.
\begin{eqnarray}
	\bar r=\frac{(\pi/2)\int_{0}^{R-L/2}r^3dr+\int_{R-L/2}^{R}r^3\arcsin(2(R-r)/L)dr}{R^2\left[(\pi/2)\int_{0}^{R-L/2}rdr+\int_{R-L/2}^{R}r\arcsin(2(R-r)/L)dr\right]}.
\end{eqnarray} 
Therefore for a rod in strong confinement, $R\rightarrow 0$, $\bar{r}=0.30$ and in weak confinement , $R\rightarrow\infty$, $\bar{r}=0.50$ as quoted in the main text.\\
For a rigid rod confined to a slit we find 
\begin{eqnarray}
	\bar r=\frac{\int_{0}^{R}r^2\arcsin(2(R-r)/L)dr}{R^2\int_{0}^{R}\arcsin(2(R-r)/L)dr}
\end{eqnarray}
for strong confinement ($L>2R$) and 
\begin{eqnarray}
	\bar r=\frac{(\pi/2)\int_{-R+L/2}^{R-L/2}r^2dr+2\int_{R-L/2}^{R}r^2\arcsin(2(R-r)/L)dr}{R^2\left[(\pi/2)\int_{-R+L/2}^{R-L/2}rdr+2\int_{R-L/2}^{R}\arcsin(2(R-r)/L)dr\right]}
\end{eqnarray} 
for ($L\le 2R$). The rigid rod in a slit should therefore have $\bar{r}=0.33$ in the limit as $R\rightarrow\infty$ and $\bar{r}=1/6$ as $R\rightarrow0$.

\begin{figure}[htbp]
	\begin{center}
		\includegraphics[width=\textwidth]{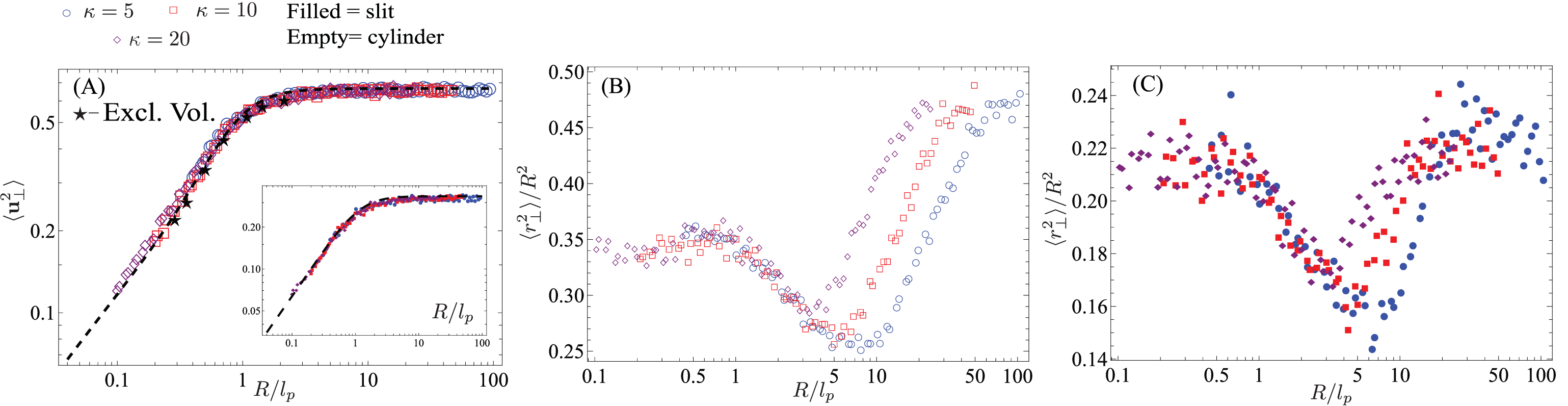}
		\caption{N=250 recreation of fig. \ref{uperpFig.fig} from the main text. There are only negligible differences between $\bar{u}$ for N=250 and N=100, mostly due to the smaller data set generated for N=250. The differences between the two for $\bar{r}$ as $R\rightarrow \infty$ are due to the change in chain length as discussed in SI sec. 5.3.}
		\label{perp250.fig}
	\end{center}
\end{figure}

\subsection{Confined Gaussian Chains}

The propagator for a Gaussian chain confined to a slit centered at the origin and of width $2R$ can be determined exactly\cite{Cordeiro:1997fu}, with 
\begin{eqnarray}
G_L(z|z_0)\propto \sum_{k=1}^\infty \sin\bigg(k\pi \frac{ (z-R)}{2R}\bigg)\sin\bigg(k\pi\frac{ (z_0-R)}{2R}\bigg)e^{-La\pi^2k^2/24R^2}
\end{eqnarray}
with an average monomer position far from the endpoints given by
\begin{eqnarray}
\langle z^2\rangle=\int_{-R}^R dzdz_0 z^2G_L(z|z_0)\bigg/\int_{-R}^R dzdz_0 G_L(z|z_0)
\end{eqnarray}.  
In the limit of $R\to 0$, this is dominated by the ground state and only the $k=1$ term need be included.  In this limit, it is straightforward to find $\langle z^2\rangle/R^2=\bar r\to 1-8/\pi^2\approx 0.19$.   For nonzero $R$ the mean can still be readily evaluated, with
\begin{eqnarray}
\langle z^2\rangle=R^2\bigg(1-\frac{8}{\pi^2}\frac{\sum_{k\ odd} k^{-4} e^{-La\pi^2k^2/24R^2}}{\sum_{k\ odd} k^{-2} e^{-La\pi^2k^2/24R^2}}\bigg)
\end{eqnarray}
Note that this has the expected limiting behavior of $\langle z^2\rangle/R^2\to 1/3$ as $R\to\infty$.  

The Gaussian propagator for cylindrically confined chains is also known\cite{Morrison:2005gz}, with
\begin{eqnarray}
G_\perp(\rv_\perp|\rv_\perp^0)\propto \sum_m\cos[m(\phi-\phi_0)]\sum_{n=0}^\infty\frac{J_m(\alpha_{mn}\rho_0/R)J_m(\alpha_{mn}\rho/R)}{J_{m+1}(\alpha_{mn})J_{m+1}(\alpha_{mn})}e^{-\alpha_{mn}^2La/6R^2}
\end{eqnarray}
where $\alpha_{mn}$ is the $n^{th}$ root of the $m^{th}$ Bessel function.  This is clearly a more tedious propagator to work with, but it can be shown that the qualitative features of $\langle \rv_\perp^2\rangle$ will remain the same:  for $R= 0$ the mean $\langle \rv_\perp^2\rangle/R^2=\bar r=(2J_2(\alpha_{00})-\alpha_{00}J_{3}(\alpha_{00}))/\alpha_{00}J_1(\alpha_{00})\approx 0.35$, and will monotonically increase to the $R\to\infty$ limit of $\bar r=1/2$.  Note that the value of $\bar r=0.35$ is somewhat larger than the minimum seen in the simulation data in the inset of Fig. \ref{uperpFig.fig}(B), consistent with the results observed in the main panel of Fig. \ref{uperpFig.fig}(B).

\end{document}